\documentclass[lettersize,journal]{IEEEtran}
\usepackage{amsmath,amsfonts}
\usepackage{algorithmic}
\usepackage{algorithm}
\usepackage{array}
\usepackage{textcomp}
\usepackage{stfloats}
\usepackage{url}
\usepackage{verbatim}
\usepackage{graphicx}
\usepackage{cite}
\usepackage{soul}
\usepackage{xcolor}
\usepackage{caption}
\usepackage{subcaption}
\begin{document}

\title{An Analytical and Empirical Investigation of Tag Partitioning for Energy-Efficient Reliable Cache}

%\author[label1]{Elham Cheshmikhani\corref{cor1}}
%%\textsuperscript{\ref{corresponding author}}}
%\author[label2]{Hamed Farbeh}
%%		
%\address[label1]{Department of Computer Science and Engineering, Shahid Beheshti University, Tehran, Iran}
%
%\address[label2]{Department of Computer Engineering, Amirkabir University of Technology (Tehran Polytechnic), Tehran, Iran}
%
%\cortext[cor1]{Corresponding author\\
%				Email address: e\_cheshmikhani@sbu.ac.ir (Elham Cheshmikhani)}

\author{Elham Cheshmikhani and Hamed Farbeh~\IEEEmembership{Member,~IEEE}
      %  \author{Elham Cheshmikhani, and Hamed Farbeh,~\IEEEmembership{Member,~IEEE}
        % <-this % stops a space
\thanks{E. Cheshmikhani (corresponding author) is with the Department of Computer Science and Engineering, Shahid Beheshti University, Tehran, Iran. E-mail address: e\_cheshmikhani@sbu.ac.ir.\\
H. Farbeh is with the Department of Computer Engineering, Amirkabir University of Technology, Tehran, Iran. E-mail: farbeh@aut.ac.ir.
}
\thanks{Manuscript received August 04, 2024; revised June 07, 2025.}}

\markboth{IEEE Transactions on Dependable and Secure Computing,~Vol.~XX, No.~X, June~2025}%
{Shell \MakeLowercase{\textit{et al.}}: A Sample Article Using IEEEtran.cls for IEEE Journals}

%%%%%%\IEEEpubid{0000--0000/00\$00.00~\copyright~2021 IEEE}

\maketitle

\begin{abstract}
Associative cache memory plays a decisive role in enhancing the performance and energy consumption of modern processors. 
Meanwhile, by occupying more than half of the processor chip area, cache memory is susceptible to transient and permanent faults, threatening the system's dependability.
As the only \textit{hardware-managed} memory module in the system, the tag array of the caches is the most critical and active component contributing a large fraction of energy consumption and error occurrence.
\textit{Tag~partitioning} is a widespread approach for both tag energy consumption reduction and reliability enhancement.
This approach splits the tag comparison operation into two steps, and only the tags whose \textit{k} lower order bits are matched with that of the input address in the first step are activated for comparing their remaining higher order bits in the second step.
The key decision parameter for tag partitioning is properly adjusting the tag-splitting point (\textit{k}) to achieve the maximum reduction in the number of reads.
This parameter has been intuitively, randomly, or experimentally selected in the existing studies without any justification.
Even for an appropriate selection of this parameter via extensive experiments, its sensitivity to various cache configuration parameters makes it ad-hoc and not extendable to other scenarios.
In this paper, we analytically illustrate that selecting an inappropriately large or small value for the tag-splitting point significantly downgrades the efficiency of tag partitioning and then formulate this parameter to determine its optimum value.
As a function of %affecting 
cache configuration parameters, the proposed formulation is proven to be convex and differentiable for determining the optimum splitting point, besides its ability to accurately report the degree of the tag partitioning efficiency for any splitting point and configuration parameters.
To approve the correctness and accuracy of the proposed formulation, we experimentally investigate the tag partitioning efficiency and optimum splitting point for a wide range of cache configurations and demonstrate a very close matching between the two. 
%%Shrink
%We also study the sensitivity of the optimum splitting point to various cache configuration parameters and depict that it gets greater in higher cache associativity, smaller cache size, and longer addresses.
%Instead of performing time-consuming exhaustive design space exploration to find a suitable tag-splitting point, 
The proposed formulation is a guarantee for the designers and researchers to instantaneously determine the optimum tag-splitting point and calculate the read reduction of tag partitioning.
\end{abstract}

\begin{IEEEkeywords}
Cache Memory, Energy Consumption, Error Rate, Tag Array, Tag Partitioning.
\end{IEEEkeywords}

\section{Introduction}
On-chip caches have a crucial role in improving system performance in modern processors that support further requirements such as Artificial Intelligence (AI) and Neural Networks (NNs) in today's life \cite{peccerillo2023ixiam, barbhuiya2023cache, cheshmikhani2022general}. 
Besides its benefit in performance and total energy consumption, on-chip caches are the most power-hungry and error-prone chip components \cite{mavropoulos2024improving}.
Among different cache structures, associative caches are preferred, while the parallel accesses to the entries in a set for tag comparisons exacerbate the cache energy consumption and error rate \cite{kim2019segmented, rostami2019parloom, farbehTC,9097450}. %%%%9492756 
The cache energy consumption further increases due to the necessity of employing Error Correction Codes (ECCs) for reliability enhancement \cite{10045828, 10328929, ghaemi2018smartag, gao2024rebec, farbehTC2}.
According to recent industrial and research reports, cache hierarchy accounts for 12-45\% of energy consumption in a system \cite{kim2019segmented}. 
%%Shrink
%Applied to today's data-dependent applications and workloads such as AI, big data, and cloud computing, this value is rapidly increasing.

%In a conventional set-associative cache, tag and data arrays are simultaneously accessed to achieve better performance, but doing so causes a large amount of energy consumption~\cite{farbehTC, ghaemi2019sleepy, farbehTC2}. 
In a conventional set-associative cache, all the target set entries in the tag arrays are simultaneously accessed to achieve better performance, but doing so causes a large amount of energy consumption~\cite{3rset, 9097450, cheshmikhani2019enhancing}. 
Among all steps of the caching procedure, tag comparison in highly-associative caches consumes a significant portion of energy dissipation.
Previous attempts, such as way prediction and sequential tag access for energy reduction, are still associated with substantial energy consumption due to the tag reads \cite{powell2001reducing}. %inoue1999way}.
Numerous new cache architectures, such as \textit{phased}, \textit{pseudo-set-associative}, \textit{way predicting}, \textit{reactive-associative}, \textit{way-halting}, and \textit{way-concatenating} caches are intended to reduce power and/or energy consumption \cite{min2004phased, inoue2003psas, zhang2003highly, liu1994cache, zhang2005way}. %inoue1999psas, inoue1999way,batson2001reactive,  zhang2000highly
%Still, they all impose some performance overheads.

%%\hl{
%On the other hand, those mentioned methods try to decrease the power consumption in SRAM-based caches, while today this technology is going to be replaced with some emerging memories.
On the other hand, the scalability challenge and high leakage power consumption of conventional SRAM-based caches in deep Nano-scale technology nodes resulted in extensive research on alternatives for SRAM technology in recent years \cite{10172078}. 
Among various emerging memory technologies, \textit{Spin-Transfer Torque Magnetic RAM} (STT-MRAM) is the most promising candidate for replacing the SRAM technology in on-chip caches \cite{hadizadeh2021copa, aspdac, hadizadeh2020stair, patent2}. 
The STT-MRAM technology benefits from high density, scalability, immunity to soft errors, and near-zero leakage power.
However, it faces some reliability challenges such as \textit{read disturbance}, \textit{retention failure}, and \textit{write failure} \cite{cheshmi, hadizadeh2021copa, cheshmikhani2019enhancing}.
Among the various sources of STT-MRAM unreliability, read disturbance is the dominant threat in the read-intensive tag array.
 %}
%The previous studies to reduce cache energy consumption are not applicable to STT-MRAM caches or impose an increase in those reliability challenges.

\textit{Tag~partitioning} is known as an effective approach for reducing accesses to tag bits in associative caches for energy consumption and/or error rate reduction \cite{kim2019segmented, liu1994cache, park2011novel, juan1996difference, fagin1997partial, petrov2001data, park2012multistep, rostami2019parloom}.
In this approach, tag comparison is split into two steps.
In the first step, $k$ lower order tag bits of the target cache set are partially compared with the requested address, and the mismatched tags are eliminated from the second step comparison in which the remaining bits are compared.  
Several recent studies in both SRAM and STT-MRAM caches that targeted tag access reduction are based on this approach \cite{3rset, kim2019segmented, liu1994cache, park2011novel, juan1996difference, fagin1997partial, petrov2001data, park2012multistep, rostami2019parloom}. 
The efficiency of tag partitioning for read reduction depends on the $tag$-$splitting~point$, where very small values for this point cause a majority of tags to be unfavorably matched in the first step, and very large values lead to a low effect of eliminating tags for the second step.

Despite its critical effect, there is no investigation on finding the proper tag-splitting point or a study on the sensitivity of both tag partitioning to this value and the proper selection of this value to the cache configuration. 
All the existing methods utilizing tag partitioning adjust tag-splitting points intuitively or experimentally for their specific system configuration. 
Its experimental adjustment requires time-consuming, exhaustive simulations to find a suitable value, and there is no guarantee of the suitability of this intuitive adjustment.

In this paper, we mathematically derive the optimum tag-splitting point for a partitioned tag structure by formulating the number of tag reads via a novel probabilistic model. 
The proposed formulation illustrates the dependency of tag reads on system configuration parameters, including cache associativity, cache size, and address length.
It is also capable of deriving the total number of tag reads per access as well as the percentage of reduction for a given splitting point, besides calculating the optimum splitting point.
As a multi-variable function, the proposed formulation is differentiable with a positive second-order derivative and is proven to be convex with a single minimum at the optimum splitting point.

We analytically and experimentally conduct a comprehensive design space exploration and demonstrate that the formulations agree with the experiments in all cache configuration scenarios.
%%Shrink
%The observations depict that the optimum splitting point slightly increases in smaller cache sizes as well as higher associativities and address lengths.
%%Shrink
%Considering today's cache configurations ranging their size from 256KB to 16MB, associativity from 4 to 32, and address length from 32 to 64 bits, our formulations and experiments reveal that the optimum tag-splitting point resides between 3 and 5. 
Taking advantage of the proposed formulation, the designers and researchers are able to not only determine the optimum tag-splitting point but also estimate the tag partitioning efficiency for energy consumption and error rate reduction.
Therefore, instead of performing time-consuming exhaustive design space exploration to find an ad-hoc suitable tag-splitting point, the proposed formulation provides an early design stage investigation.

The rest of this paper is organized as follows. 
In Section~\ref{sec:two}, we explore the related studies and present motivation for this work. 
In Section~\ref{sec:analytical}, we propose different formulations to analyze the effective parameters on the optimal number of read bits in partial tag comparison. %and derive the optimum splitting point. 
The analytical investigations and comprehensive empirical evaluations are presented in Section \ref{sec:eval}. 
We conclude the paper in Section~\ref{sec:conclusion}.

 \begin{figure}[t]
%\captionsetup{font=footnotesize}
	\centering
		\includegraphics[width=0.6\linewidth]{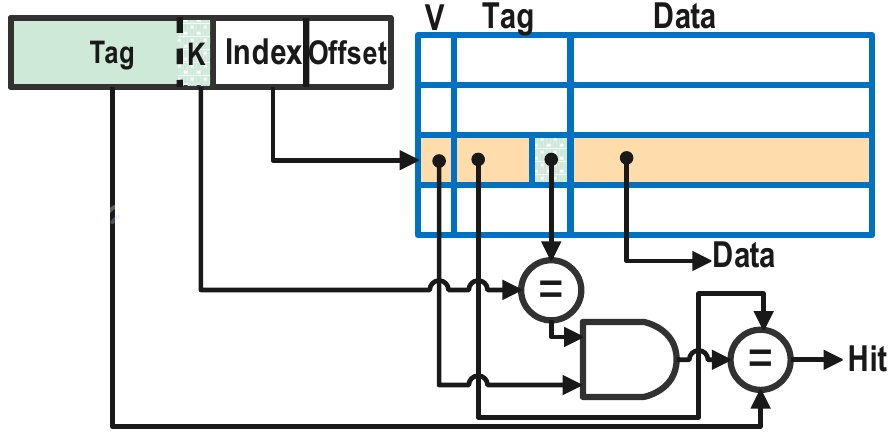}%Fig11.jpg
	\caption{Two-step tag comparison in tag partitioning method.}\vspace{-10pt}		
	\label{fig:partial}\vspace{-5pt}		
\end{figure}

%\vspace{-10pt}
\section{Related Work and Motivation}\label{sec:two}
Several methods have been proposed to manage the tag bits comparison in SRAM and STT-MRAM caches \cite{kim2019segmented, liu1994cache, park2011novel, juan1996difference, fagin1997partial, petrov2001data, park2012multistep, rostami2019parloom, 3rset, patent1}.
These methods are motivated by the observation that mismatches in tag comparisons typically take place in the low order bits due to the spatial locality.
Therefore, a tag partitioning approach for partial tag comparison is proposed to reduce power consumption and decrease the number of read bits in the cache.
Fig. \ref{fig:partial} depicts an abstract schematic of tag partitioning and its two-step comparison operation.
%Cache memory systems generally consume 12-45\% of processor energy \cite{sodani2011race}.

\subsection{Tag Partitioning Based on Way Prediction}
The partial tag comparison was originally proposed for way-prediction \cite{liu1994cache, juan1996difference} to decrease the cache power consumption.
The 2-way set-associative cache in these works played a direct-map cache role to gain power saving.
The cache way is selected based on noticing that the two tags for a set must differ by at least one bit.
These studies are the basis of all partial tag comparisons for power-saving goals.

Another study tries to predict cache misses using a Bloom filter \cite{park2012multistep}.
In this work, each of the cache ways is fed to a Bloom filter.
They use the $Most$-$Recently~Used$ (MRU) policy based on the $hot$ lines to predict the cache hits, and if the numerical values for hot lines are small amounts, a Bloom filter is employed to predict the cache misses.
%%Shrink
%It is mentioned that lower order 3 bits have been used for partial tag comparison with no justification for this selection.
An enhanced counting Bloom filter is also used in \cite{rostami2019parloom} to predict cache misses with high accuracy, as well as partial tags to decrease the overall cache size.
This way, unnecessary tag comparisons can be prevented, and therefore, the cache energy consumption is considerably reduced.
%%Shrink
%This method is only applicable to the L1 cache.

The scheme proposed in~\cite{park2011novel} for tag comparison reduction is based on predicting either cache hits or cache misses.
The authors presented a partial tag-enhanced Bloom filter to improve the accuracy of the cache miss prediction and hot/cold checks that control data liveness to reduce the tag comparisons of the cache hit prediction method.
In this study, the number of lower order bits to be compared with the input tag address can be changed based on the blocks that are grouped between hot and cold blocks and the number of their accesses.

\vspace{-5pt}
\subsection{Tag Partitioning via Buffering Lower Order Bits}
A partial tag comparison is proposed for SRAM-based caches in \cite{min2004partial}.
The authors added a very small array of tags 
%%Shrink
%(3  least significant bits of the original tag array) 
beside the original tag array.
These added tags are compared with the target address. 
%%Shrink
%The number of tag bits chosen for comparison is 3 and there is a discussion that if they make it larger, there is no difference in performance improvement.
%Therefore, the authors choose to compare 3 lower bits in the first step.
If none of the ways' 3 least significant bits match the income address, a miss is reported.
Only when there is a match, the sense amplifiers attached to the data array bit-lines will be enabled to read the corresponding data.

In \cite{chen2005low}, the authors extended the tag array with an additional register file, which consists of 2 lower order bits of each tag array way.
The comparison procedure contains three steps. 
First, the 2 lower order bits in the register file are compared with the input tag address.
%%Shrink
%If any match cannot be found in this step, a miss is reported.
If one or more matches occur by comparing those 2 bits in the first step, they have to compare the other bits, as well.
If the remained bits are matched with the input tag address, the corresponding data block is read from the data array in the third step. 
%%Shrink
%It should be noted that there is no point in the source of 2-bit selection in this study.

One of the most popular studies in SRAM-based caches to decrease power dissipation is the phased tag cache \cite{min2004phased}. 
This method tries to partially compare the tags.
The 7 least significant bits
%%Shrink
%, which are called identity bits, 
are compared in the first phase, and the remaining bits are compared in the second phase.
%In this study, the chosen bits in the first phase are selected as 3 bits or 7 bits based on the cache type.

A new cache architecture, called \textit{way-halting~cache}, is developed in \cite{zhang2005way} to reduce the energy consumption.
Way-halting cache is a four-way set-associative cache that stores the four lowest-order bits of all ways' tags into a fully associative memory, called a halt tag array.
%The lookup in the halt tag array is done in parallel with and is no slower than, the set-index decoding.
The halt tag array predetermines which tags cannot match due to their low order 4 bits mismatching.
Further accesses to ways with known mismatching tags are then halted, thus saving energy. 
%%Shrink
%The authors discussed that the purpose of selecting 4 bits for comparison is that it comes closest to the ideal of accessing one way for each hit and zero ways for each miss. 

Compressed tag architecture is introduced in \cite{kwak2010compressed}.
To reduce the energy consumption in tag arrays, this tag-matching mechanism uses a locality buffer and tag compression.
Small tag space matching is used instead of full tag matching in this work.
Initially, the small tag space %%Shrink
%consists of two bits, 
and then the remaining bits are compared to be matched.
%The number of bits to be read in the small tag space was randomly selected.

\subsection{Tag Partitioning by Splitting Tag}
In \cite{kim2019segmented}, a partial tag comparison method called \textit{Segmented Tag Cache} (STC) is proposed to reduce the amount of energy consumption during the tag access.
%The proposed method is based on partial tag access. 
In this method, the tag array is segmented into two parts, and partial tag access only reads the low order part of the tag. 
%Under the scheme, the delay of the tag organization is hidden by the data access delay, avoiding an increase in the overall cache access time to be maintained.
%Similar to way halting cache technique, the STC is designed to access the low-order bits more quickly than the normal tag access speed.
The STC modifies the tag array to reduce its overhead same way as the way-halting cache.

Reduced One-Bit Tag Instruction Cache (ROBTIC) was proposed in \cite{gu2011chip} that enables the cache tag field to only contain the least significant bit of the full tag in the first step.
As a result, the cache can only be mapped to one segment of memory, whereas it covers two regions.
Next, the most significant bits of the memory addresses can be used to identify the mappable regions.

A new filter cache model using partial tag matching 
%%Shrink
%for energy efficiency 
named \textit{Data Filter Cache System with Partial Tag Cache} (DFPC) is proposed to reduce L1 cache access energy \cite{choi2014data}.
This model employed another small cache, called \textit{partial tag cache}, 
%%Shrink
%which is accessed simultaneously with the filter cache and used
to filter unnecessary accesses to the ways of cache.
%This method led to the reduction of energy consumption by minimizing the number of accesses to the L1 cache.
This work used 6 bits to compare the partial tag cache.
%The matching cover, which is a reduction in the number of false hits, is mentioned as the reason for this selection. 

\subsection{Tag Partitioning in STT-MRAM Caches}
In \cite{park2012future}, the authors presented two techniques -- sequential tag-data access for reads and partial line update for writes -- that improve the energy efficiency of STT-MRAM caches.
A cache architecture that performs partial cache line updates for cache writeback energy reduction was proposed.
The technique does not incur any extra cache misses since it does not change the data flow between different cache levels.

Another study tried to increase the reliability of STT-MRAM caches using partial tag comparison \cite{3rset, patent1}.
The authors split the tag comparison into two parts, first, they read the 4 bits of low order bits and in the next step, the remained bits are read and compared with the input address.
They experimentally demonstrated that among different splitting points, 4 represents better reliability and energy saving for their dedicated cache configuration.

%
%A partial resolution method was proposed in data value predictors \cite{sato2000partial}.
%By using more tag address bits than necessary, this study examined the possibility of reducing hardware budgets for data value predictors.
%They showed that two tag bits are enough in the first step of tag comparison.
%The key contribution of this work is the introduction of various partial tag resolution policies, such as stride, last-value, two-level, etc. \cite{kwak2010compressed}.

All the tag partitioning methods require a two-step operation for tag comparison.
In the first step, a part of the tag bits are read and compared with the associated bits, and in the second step, the remained bits of a subset of tags are read and compared.
All the methods dedicate a number of bits for the first step based on the reduction of false hits or increasing the number of hits.
By the way, changing the cache size, the application running on the cache, the address length, or the cache associativity can affect this number.
None of those studies provides an analytical reasoning for their selection, while this value can have a significant effect on their efficiency for reducing energy consumption or error rate.
In this paper, we open a new door to analyze the affecting factors of this selection.
In this case, given the size of the cache, associativity, and other parameters of the cache, the optimal number of bits that should be read in the first step of partial tag comparison is determined to maximize the efficiency of the tag partitioning-based methods.

%Most of the solutions to decrease the power consumption in cache memories based on partially tag comparisons usually require complex cache arrangements such as circuitry for selectively activating or deactivating columns of the cell array.
%On the other hand, those methods are applicable to the L1 cache, while they are not applicable to the last level cache due to the less reading operation in the last level caches.
%Simultaneously, the  previous studies are based on SRAM technology, while for emerging memories these methods cause some challenges if they are applicable to those emerging memories.

\section{Proposed Analysis and Formulation}\label{sec:analytical}
Partially comparing tags requires reading a few low order bits in the first step and the remaining part of the matched tags in the next step.
In order to dedicate the proper number of reads in each step to minimize the number of reads for achieving the best reliability and energy saving, different parameters must be considered.
In this paper, we formulate the number of tag reads to mathematically determine the tag-splitting point for maximum efficiency.
It is important to note that tag partitioning does not influence cache access patterns, cache misses, interactions between cache levels, or operations related to coherency.
The sole aspect impacted by tag partitioning is the frequency of reads from the tag cell, which is altered by modifying the tag comparison circuitry.

%In addition to the partial tag comparison's aim of reducing the energy consumption of both SRAM and STT-MRAM caches, it increases the reliability of STT-MRAM caches.
%These goals are achieved by reducing the number of bits that need to be read and compared.
%Consequently, we must propose a mathematical equation that calculates the total number of read bits in a cache and then try to minimize it.

     \begin{figure*}[t]
%\captionsetup{font=footnotesize}
	\centering%\vspace{-15pt}
		\includegraphics[width=0.9\linewidth]{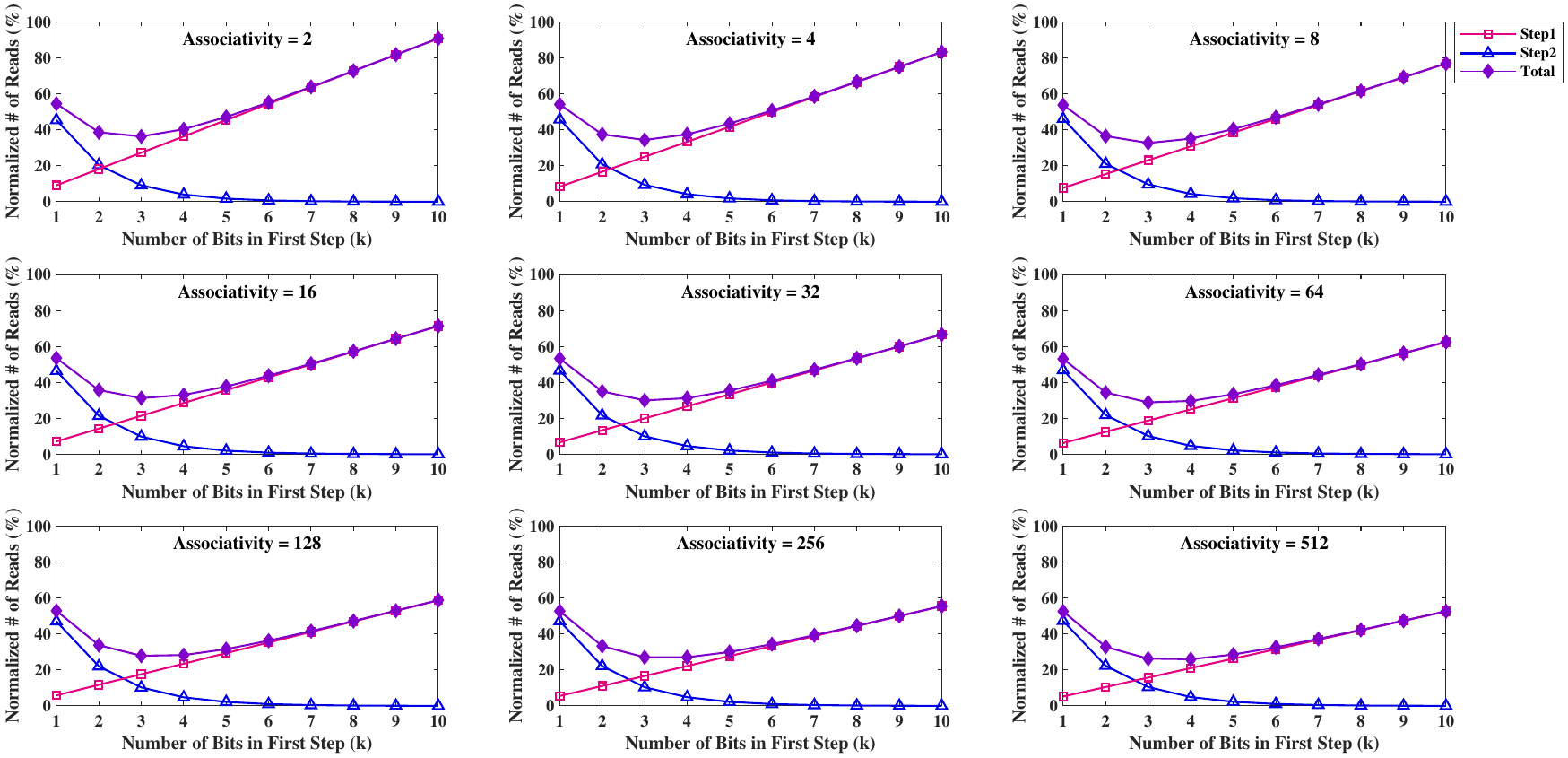}
	\caption{Expected number of reads in the first (Eq. (\ref{eq:1a})) and second steps (Eq. (\ref{eq:1s})) besides their accumulation based on Eq. (\ref{eq:1}).}	\vspace{-5pt}		
	\label{fig:workload}
\end{figure*}

\subsection{Formulating Number of Tag Reads}
Considering an ${x}$-way set-associative cache, the number of reads from tag bits in conventional cache is according to (\ref{eq:0a}).
 \vspace{-5pt}
 \begin{equation}
  \begin{split}
  \label{eq:0a}%\hspace{-2.9cm}
  Number~of~read~bits~in~baseline = n \times x 
%  _{intrinsic}) + (error~rate_{temperature-induced})
  \end{split}
  \end{equation}
where, \textit{n} is tag length, which depends on address length, cache size, associativity, and cache block size.
In the partitioned tag, the number of reads in the first step is given in (\ref{eq:1a}).
 \begin{equation} \vspace{-5pt}
  \begin{split}
  \label{eq:1a}%\hspace{-2.9cm}
  Number~of~read~bits~in~first~step = k \times x 
%  _{intrinsic}) + (error~rate_{temperature-induced})
  \end{split}
  \end{equation}
where, \textit{k} is the number of bits compared in the first step, i.e., the tag-splitting point.
The number of bits \textit{per tag} to be compared in the second step is as simple as (\ref{eq:1b}).
  \begin{equation} \vspace{-5pt}
  \begin{split}
  \label{eq:1b}%\hspace{-2.9cm}
  Number~of~read~bits~in~second~step =  n~-~k
%  _{intrinsic}) + (error~rate_{temperature-induced})
  \end{split}
  \end{equation}  
where, \textit{n} is the total tag length.
However, calculating the total number of reads in the second step is more complicated, as it requires determining the number of tag ways matched in the first step.
We model this value as a probability problem. Considering \textit{k} bits in partial tag comparison of the first step, the probability of a match is according to (\ref{eq:1c}).
\begin{equation} \vspace{-5pt}
  \begin{split}
  \label{eq:1c}%\hspace{-2.9cm}
  Probability~of~a~match~in~first~step = 
   \frac{1}{2^k}
%  _{intrinsic}) + (error~rate_{temperature-induced})
  \end{split}
  \end{equation}

Equation (\ref{eq:1c}) is equivalent to the probability of success in a single Bernoulli trial. 
Equivalently, the probability of a mismatch is according to (\ref{eq:1d}).
	\begin{equation}
  \begin{split}
  \label{eq:1d}%\hspace{-2.9cm}
  Probability~of~a~mismatch~in~first~step = 
 1-\frac{1}{2^{k}}
%  _{intrinsic}) + (error~rate_{temperature-induced})
  \end{split}
  \end{equation} 
Consequently, the number of matched tags in the first step for an x-way set-associative cache can be calculated using (\ref{eq:1e}).
\begin{equation}
  \begin{split}
  \label{eq:1e}%\hspace{-2.9cm}
  Number~of~matched~tags~in~first~step = \\
  \sum_{i=0}^x i\times \binom{x}{i} \times (\frac{1}{2^k})^i \times (1-\frac{1}{2^{k}})^{x-i}
  \end{split}
  \end{equation}
 Combining (\ref{eq:1b}) and (\ref{eq:1e}) gives the number of reads in the second step, as shown in  (\ref{eq:1s}).    
\begin{equation}\vspace{-5pt}
  \begin{split}
  \label{eq:1s}%\hspace{-2.9cm}
  Number~of~read~bits~in~second~step = (n~-~k)\times \\
    \sum_{i=0}^x i\times \binom{x}{i} \times (\frac{1}{2^k})^i \times (1-\frac{1}{2^{k}})^{x-i}
%  _{intrinsic}) + (error~rate_{temperature-induced})
  \end{split}
  \end{equation}

The summation of (\ref{eq:1a}) and (\ref{eq:1s}) results in the total number of reads for a cache access request as given in (\ref{eq:1}).    
\begin{equation}\vspace{-5pt}
  \begin{split}
  \label{eq:1}%\hspace{-2.9cm}
  Total~number~of~expected~read~bits = k \times x + (n~-~k)\times \\
    \sum_{i=0}^x i\times \binom{x}{i} \times (\frac{1}{2^k})^i \times (1-\frac{1}{2^{k}})^{x-i}
%  _{intrinsic}) + (error~rate_{temperature-induced})
  \end{split}
  \end{equation}
where, \textit{n} is the number of tag address bits, \textit{k} is the number of read bits in the first step, and \textit{i} is the number of tag matches in the first step.
%It is noteworthy that \textit{i} can get a value between ``0'' and ``\textit{x}''.

\subsection{Analysing Tag Splitting Effect}
As shown, this equation has two parts; the first part is ${k \times x}$, which is the number of read bits in the first step.
It demonstrates that \textit{k} bits, which is the tag-splitting point, are read in the first step from all the ways of the tag array.
For the second step, the remaining (\texttt{n} $-$ \texttt{k}) bits of the mismatched tags are discarded.
The number of matched ways compared in the second step is calculated based on the given probability.

Considering the effect of the splitting point on the number of reads in each step, (\ref{eq:1a}) shows that the number of reads in the first step linearly increases by larger \textit{k}, whereas the number of reads in the second step super-linearly decease according to (\ref{eq:1s}). 
In other words, the two terms in the total number of reads are adversely and differently affected by changing \textit{k}. 

Fig. \ref{fig:workload} illustrates the effect of \textit{k} on the number of reads in each step as well as the summation of the two. The results are for a wide range of cache associativities from 2 to 512 in a 1MB cache size and address length of 40 bits.
The horizontal axis is the tag-splitting point, and the vertical axis is the number of reads normalized to the baseline, i.e., non-partitioned tag.
The curves related to the first and second steps show that their different ascending and descending trends lead to a minimum for the curve of total reads. The splitting point at which the minimum of total reads takes place is the optimum point we are looking for. 
Considering various associativities, the optimum splitting point increases from 2 in the 2-way cache to 4 in the 512-way cache, which indicates that the optimum value for \textit{k} is directly but very slightly affected by the cache associativity. 
Besides the above observation derived from the formulation of the number of reads, we provide a formal calculation of the optimum splitting point in the following.

\begin{table*}[t]%\vspace{-5pt}
				\centering%\vspace{-1pt}
				\caption{Number of reads using Eq. (\ref{eq:1}) for the optimal point ($k_{optimal}$ and its rounded value $k_{min}$) calculated by Eq. (\ref{eq:4}).}%\vspace{-10pt}
				\includegraphics[width=0.95\linewidth]{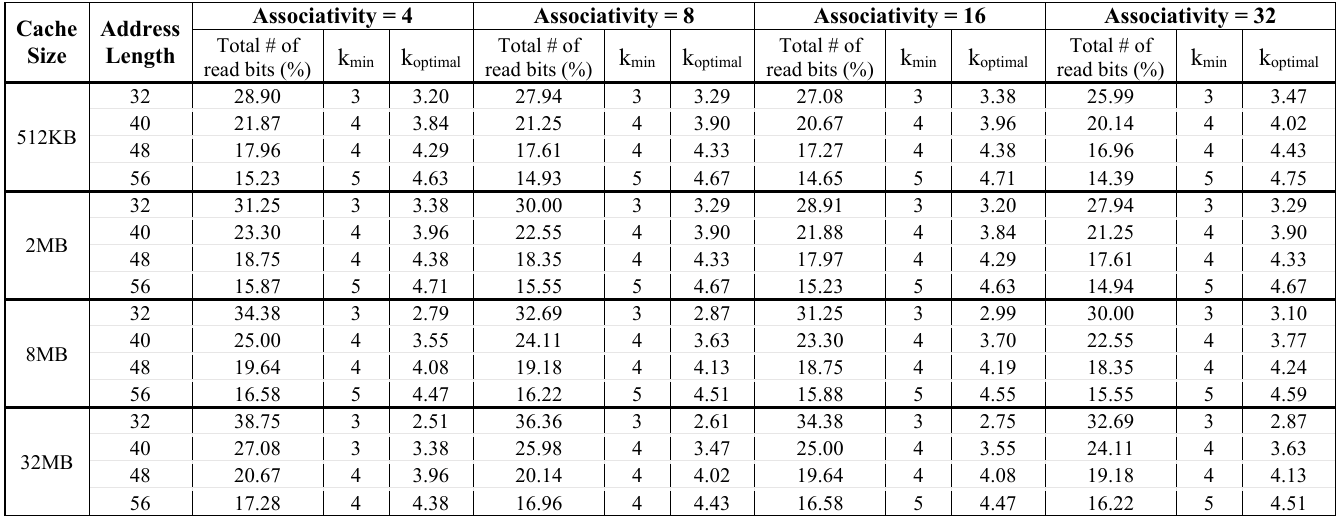}\vspace{-5pt}%table-1
				\label{table:1}%\vspace{-10pt}
\end{table*}

\subsection{Formulating Optimum Tag Splitting Point}
So far, the total number of read bits in the tag part of the cache using partial tag comparison has been formulated, as observed in (\ref{eq:1}).
%To find out the optimum number of$k$, the total number of read bits should be minimum. 
The aim of this formulation is to calculate the optimal value of \textit{k}, which is the one that minimizes the total number of read bits in both steps of partial tag comparison.
To do so, we investigate the convexity of (\ref{eq:1}) and therefore, the second derivative of the function should be calculated.
%\green{REMOVE: Since hit rate is a constant value for a given workload and the equations will be differentiated from now on, we calculate (\ref{eq:1})'s derivative for the sake of visibility.}
The second derivative of (\ref{eq:1}) is shown in (\ref{eq:2}). 

\begin{figure}[t]
%\captionsetup{font=footnotesize}
	\centering\vspace{-10pt}
		\includegraphics[width=0.8\linewidth]{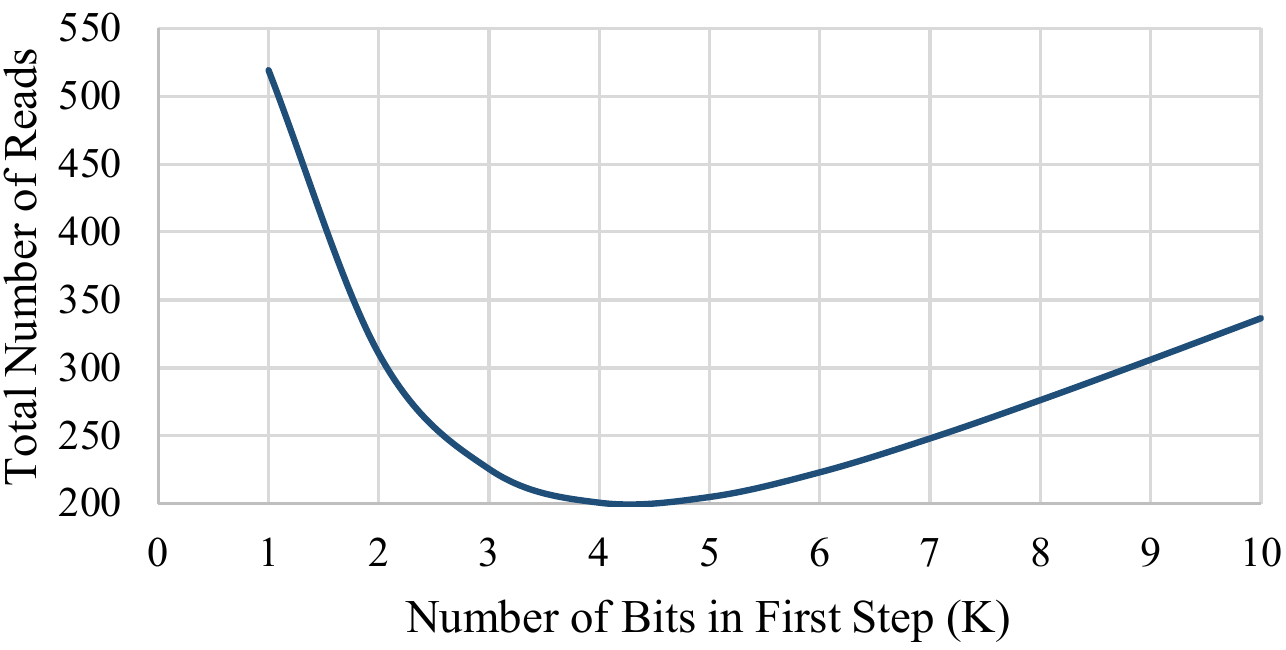}%Fig1.pdf
	\caption{Proving the convexity of Eq. (\ref{eq:1}), for 40-bit tag length and 16-way set-associative.%Elham:finally 35 or 40 or 42?
	}\vspace{-10pt}		
	\label{fig:esbat}
\end{figure}

\vspace{-5pt}
\begin{equation}
\begin{split}
 \label{eq:2}\vspace{6pt}%\hspace{-2.9cm}
 \frac { \partial^2{[k \times x + (n~-~k)\times \sum_{i=0}^x i \times \binom{x}{i} \times (\frac{1}{2^k})^i \times (1-\frac{1}{2^k})^{x-i}}]}{ \partial k}\\
{=\frac{x}{2^k}\times \log 2 \times (2+(n-k)\times \log 2)}
\end{split}
\end{equation}
Since the value of \textit{k} is between 0 and \textit{n}, the second derivative of (\ref{eq:1}) is positive definite.
Therefore, (\ref{eq:1}) is \textit{convex} as shown in Fig. \ref{fig:esbat} and has a minimum, which is the optimal number of \textit{k}.

To calculate the minimum of (\ref{eq:1}), its first derivative should be calculated and equal to zero. 
We can see the first derivative of (\ref{eq:1}) in (\ref{eq:3}).
\begin{equation}
\begin{split}
 \label{eq:3}\vspace{6pt}%\hspace{-2.9cm}
 \frac { \partial{[k \times x + (n~-~k)\times \sum_{i=0}^x i \times \binom{x}{i} \times (\frac{1}{2^k})^i \times (1-\frac{1}{2^{k}})^{x-i}}]}{ \partial k}\\
{=\frac{x}{2^k}\times (\log 2 \times (k-n) + 2^{k}-1)}
 \end{split}
 \end{equation}

The equality of (\ref{eq:3}) to zero results in (\ref{eq:4}), where \textit{ProductLog} is the Lambert $W$ function, also called the omega function or product logarithm. 
The multivalued function of product logarithm (\texttt{ProductLog[z]}) equals $w$, if $z=we^w$.
\begin{equation}
\label{eq:4}\vspace{6pt}%\hspace{-2.9cm}
k = \frac { \log (ProductLog[2^n\times e])}{\log 2}
\end{equation}

%\begin{figure*}[t]%\vspace{-20pt}
%				\centering
%				\subfloat{\includegraphics[width=1\linewidth]{surface1-HF.pdf}}%\vspace*{-6pt}1aa
%				%\hfill
%				\vspace{-2pt}
%			
%				\subfloat{\includegraphics[width=0.97\linewidth]{surface2-HF.pdf}}\vspace{-5pt}%Fig3--1bb
%				\caption{STT-MRAM read and write operations: (a) reading `0', (b) reading `1', (c) writing `0', and (d) writing `1'.}\vspace{-10pt}
%				\label{fig:basics}
%			\end{figure*}

The final goal of the above analysis is to derive the formula of the optimum splitting point, which is given in (\ref{eq:4}).
According to (\ref{eq:4}), it seems that \textit{k} is, apparently, determined only by \textit{n}.
Since \textit{n} is the tag length, both cache size and associativity directly affect it, in addition to the address length.
%\green{REMOVE: Besides, the access pattern, which determines the cache hit and miss influences this number.
%In this paper, we demonstrate which one's effect is more than the others.}

In this paper, we examine this dependency for different associativities and different cache sizes.
By increasing the cache size, the tag address length is decreased.
As illustrated in Table \ref{table:1}, the total number of read bits is calculated based on (\ref{eq:1}).
The result of the total number of read bits in comparison with the optimal number of bits to be read extracted from (\ref{eq:4}) shows that in an 8-way set-associative cache, reading \texttt{4} bits in the first step results in the minimum number of total read bits when the address length is 40 or 48.
The optimum number of read bits in the first step for 16- and 32-way set-associative caches with the same address length remains \texttt{4}.
%This optimal number of reads is the same for all the caches in different ways, as long as the address length is the same in comparing both columns of the total number of read bits and optimum \textit{k} of (\ref{eq:4}).
%This value changes to five when the cache size, address length, or associativity increases.
This value changes to \texttt{5} in very large associativities, very small caches, and/or very long addresses, and changes to \texttt{3} conversely. 
The above theoretical calculations depict that the optimum splitting point for the partial cache comparison is in the range of \texttt{[3, 5]} for conventional cache configurations.

%
%\begin{scriptsize}
%\begin{table}[t]
%\label{table:iii}
%  \centering
%  \caption{Total number of tag bits using equations (\ref{eq:1}) and (\ref{eq:4}).}
%  \label{table:formatting}
%  \begin{tabular}{|l|l|l|}
%    \hline
%    \textbf{Total~\#~of~tag~read~bits} & \textbf{$k_{minimum~\#~of~bits}$} & \textbf{$k_{optimal}$}\\
%    \hline
%    \hline
%    File format & PDF & 4\\
%    \hline
%%    Page limit & 11 pages, {\bf not including}\\
%%               & {\bf references}\\
%%    \hline
%%    Paper size & US Letter 8.5in $\times$ 11in\\
%%    \hline
%%    Top margin & 1in\\
%%    \hline
%%    Bottom margin & 1in\\
%%    \hline
%%    Left margin & 0.75in\\
%%    \hline
%%    Right margin & 0.75in\\
%%    \hline
%%    Body & 2-column, single-spaced\\
%%    \hline
%%    Space between columns & 0.25in\\
%%    \hline
%%    Line spacing (leading) & 11pt \\
%%    \hline
%%    Body font & 10pt, Times\\
%%    \hline
%%    Abstract font & 10pt, Times\\
%%    \hline
%%    Section heading font & 12pt, bold\\
%%    \hline
%%    Subsection heading font & 10pt, bold\\
%%    \hline
%%    Caption font & 9pt (minimum), bold\\
%%    \hline
%%    References & 8pt, no page limit, list \\
%%               & all authors' names\\
%   % \hline
%  \end{tabular}
%\end{table}
%\end{scriptsize}

%\begin{table}[t]%\vspace{-5pt}
%				\centering%\vspace{-1pt}
%				\caption{System Configuration.}\vspace{-10pt}
%				\includegraphics[width=0.5\linewidth]{TABLE3.pdf}\vspace{-5pt}%table-1
%				\label{table:2}\vspace{-10pt}
%\end{table}

\begin{table}[t]%\vspace{-10pt}
				\centering%\vspace{-1pt}
				\caption{Details of system configuration.}%\vspace{-10pt}
				\includegraphics[width=1\linewidth]{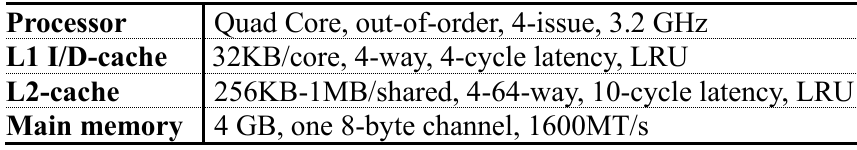}\vspace{-5pt}%table-1
				\label{config}\vspace{-10pt}
\end{table}

\section{Evaluation Methodology and Results}\label{sec:eval}
\subsection{System Setup}
This section experimentally evaluates and verifies the formulations and analysis presented in this work and investigates its various aspects.
To this aim, a quad-core processor equipped with dedicated instruction/data L1 caches and a shared L2 cache is modeled in the gem5 full-system simulator \cite{gem5}.
Table \ref{config} depicts the details of the system configuration.
The workloads are 17 combinations of randomly-selected memory-bound and CPU-bound programs in the SPEC CPU2017 benchmark suite \cite{spec}, as given in Table \ref{workload}, simulated for 4 billion instructions after fast-forwarding the initial 200 million instructions as the warm-up phase.

\begin{table}[t]\vspace{-5pt}
				\centering%\vspace{-1pt}
				\caption{Combinations of benchmarks for generating workloads for a simulated quad-core processor.}\vspace{-5pt}
				\includegraphics[width=0.9\linewidth]{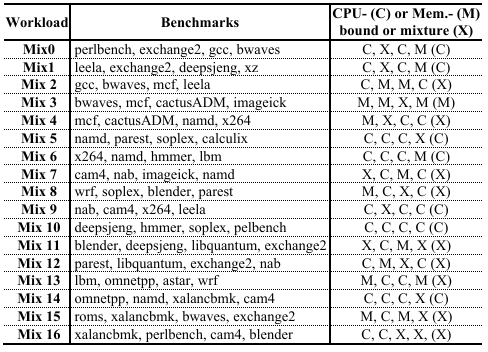}\vspace{-5pt}%table-1
				\label{workload}\vspace{-10pt}
\end{table}

As mentioned in Section \ref{sec:analytical}, the key parameters affecting the efficiency of the tag partitioning approach are the degree of cache associativity, address length, and cache size.
Cache partitioning can be utilized in different cache levels, and the goal of this study is to show that the optimum splitting point can be formulated and calculated in advance with no requirement for workload profiling. 
All the previous theoretical observations, i.e., Fig. \ref{fig:esbat}, Fig.\ref{fig:workload}, and Table \ref{table:1}, are independent of cache level and are applicable to both L2 and L3 caches as well as L1 cache, considering their suitable configurations.
To investigate the conformity of the formulations with the experiments, tag splitting is applied to the L2 cache in the following evaluations. 
Three cache sizes, i.e., 256KB, 512KB, and 1MB, have been considered in the experiments to correspond to the L2 cache size in real systems. In addition to regular cache associativities (8-way and 16-way), we also evaluated very small and very large associativities, e.g., 4-, 32-, and 64-way, in L2 cache to illustrate the matching between the experiments and formulation even in extreme configurations and corner cases.

We use CACTI to model the SRAM cache in various configurations within a 32nm technology node, and the energy parameters are detailed in Table \ref{energy} for a 256KB L2 cache with 8- and 16-way associativity. For the sake of brevity, we are unable to report the energy details for all evaluated cache sizes and configurations.
Since tag partitioning has no impact on the number of cache memory cells or cache performance, and the modifications to the cache control circuitry are negligible, the leakage power/energy in conventional and tag-partitioned caches is nearly identical.
Targeting to reduce read operations from the tag array, tag partitioning primarily influences dynamic energy consumption. The three highlighted rows in Table \ref{energy} represent the tag energy parameters impacted by tag partitioning. As shown, while certain cache components remain unchanged in tag partitioning, these three highlighted rows account for the majority of total tag energy consumption.

 \begin{table*}[t]%\vspace{-5pt}
				\centering%\vspace{-1pt}
				\caption{Details of energy parameters in the tag array for a 256 KB cache considering various address lengths.}\vspace{-5pt}
				\includegraphics[width=0.95\linewidth]{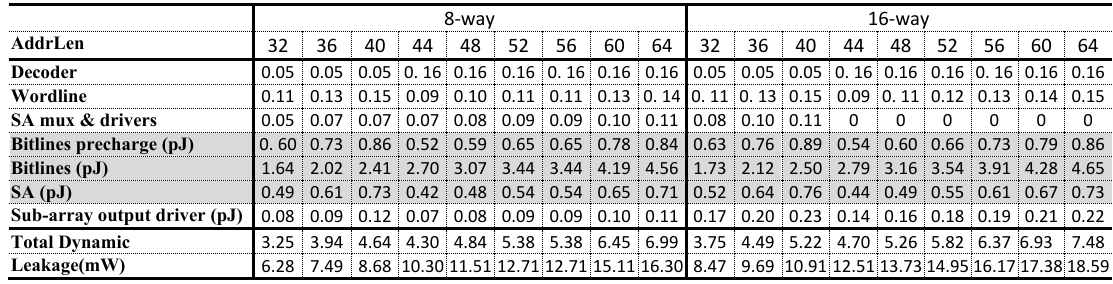}\vspace{-5pt}%table-1
				\label{energy}%\vspace{-10pt}
\end{table*}

 \begin{figure*}[h]
%\captionsetup{font=footnotesize}
	\centering
		\includegraphics[width=0.9\linewidth]{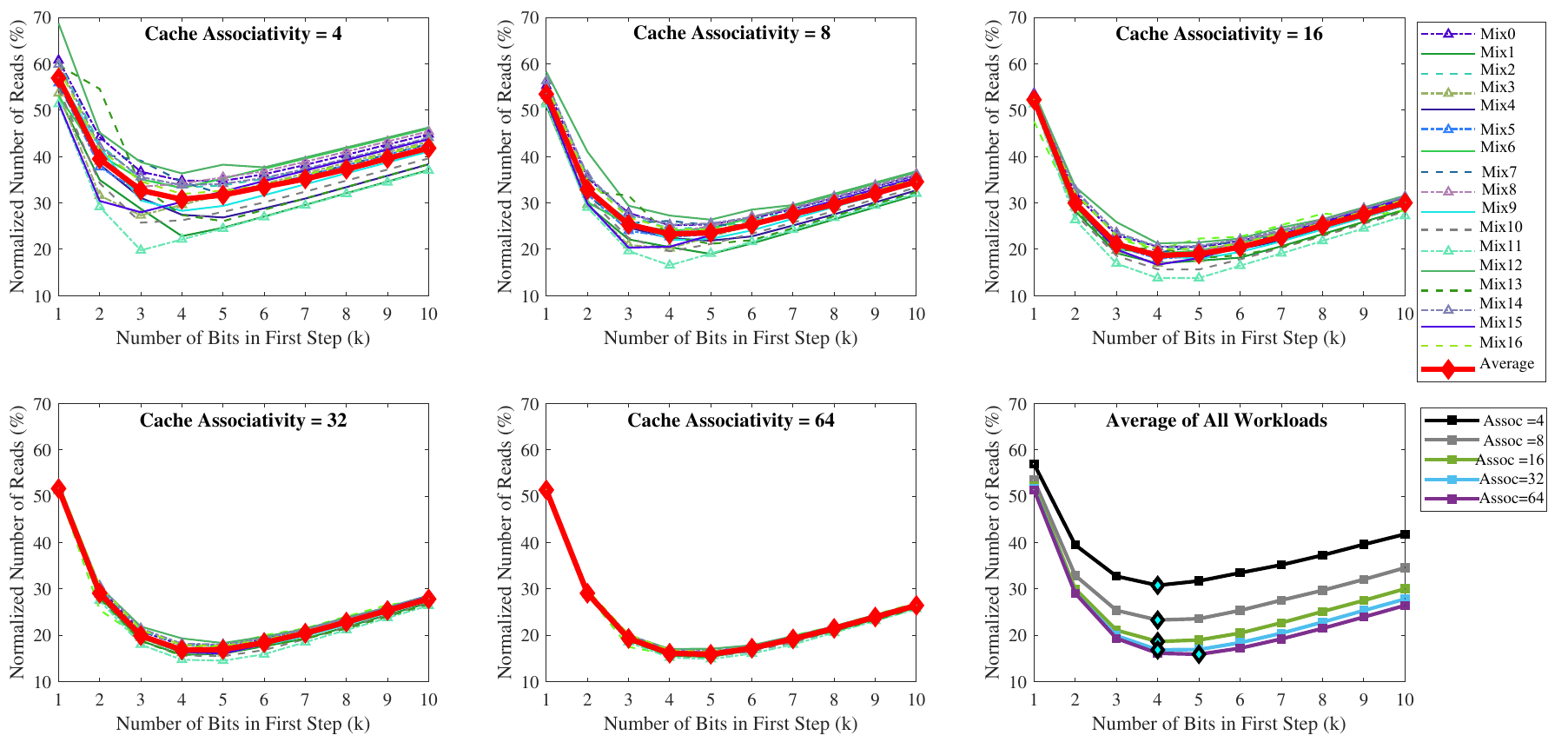}%2017
	\caption{The number of tag reads as a function of splitting points for various cache associativity in all workloads.}\vspace{-5pt}		
	\label{fig:workload-main}\vspace{-5pt}	
\end{figure*}

For a comprehensive design space exploration, a wide range of cache configurations is explored.
% including cache associativity from 2 to 64, address length from 40-bit to 64-bit, cache size from 256KB to 8MB, and tag-splitting point from 1 to 10 are evaluated.
The evaluations are for tag-splitting points from 1 to 10. Limiting the tag-splitting point to a maximum of 10 is because both formulations and experiments illustrate that the optimum point for all configurations is always significantly lower than 10.
In the following, we first discuss the various workloads' behaviors and then illustrate the conformation of experiments and formulations. 
Then, the effect of the tag splitting point on energy consumption and error rate, and the proper selection of this value, is explored.
%Table \ref{table:II} depicts the 420 combinations of each 6 cache sizes for these configurations.

\subsection{Experimental Tag Read Evaluation}
The number of tag bits reads per cache access request is the critical evaluation parameter that affects both the tag energy consumption and error rate.
For each workload in a specific configuration, we normalize the tag bits read in each splitting point to the number of reads in the baseline, i.e., the cache configuration without partial tag read.
Fig. \ref{fig:workload-main} illustrates the number of reads in various tag-splitting points for all workloads considering cache associativity ranging from 4 to 64, cache size of 1MB, and 40-bit address length.
It is noteworthy that a similar trend is observed in the other cache sizes/address lengths, and the deductions are extendable to other configurations.

In Fig. \ref{fig:workload-main}, five cache associativities (from 4 to 64) are separately plotted in independent subfigures, and the averages of all workloads in the five cache associativities are integrated into the sixth subfigure to better illustrate the effect of associativity on the optimum splitting point.
Considering the trends in these associativities, three key observations are evident.
First, by increasing the cache associativity, the variation and diversity in the workloads are reduced, and their behavior in the number of reads in various splitting points is smoother and closer to the average curve.
This shows that although the reduction in the number of reads is workload-dependent, this dependency loosens in larger cache associativities due to a higher degree of randomness in low order tag bits in a set.
Hence, the optimum splitting point is not meaningfully affected by the workload behavior, but by the cache configuration.
Second, higher read reductions are observed in larger cache associativities.
This is because a greater number of tags in a set are eliminated in larger associativities. 
Third, the optimum splitting point for all workloads in all associativities is almost the same and resonates between 3 and 5, which confirms our formulations.

%\begin{figure*}[!h]
%    \centering
%    \begin{subfigure}[b]{0.8\textwidth}
%           \centering
%		 \includegraphics[width=0.8\linewidth]{Line1-Final.pdf}%\vspace*{-6pt}1aa
%				    \end{subfigure}
%				\vspace{-2pt}
%			\begin{subfigure}[b]{0.8\textwidth}
%           \centering
%				 \includegraphics[width=0.8\linewidth]{Line2.pdf}%\vspace{-5pt}%Fig3--1bb
%					    \end{subfigure}
%					\vspace{-1pt}
%				\begin{subfigure}[b]{0.8\textwidth}
%           \centering
%				 \includegraphics[width=0.8\linewidth]{Line3.pdf}%\vspace*{-6pt}1aa
%				    \end{subfigure}
%				\vspace{-1pt}
%			
%			\begin{subfigure}[b]{0.8\textwidth}
%           \centering
%				 \includegraphics[width=0.8\linewidth]{Line4.pdf}%\vspace{-5pt}%Fig3--1bb
%				     \end{subfigure}
%				\vspace{-1pt}
%				%\hfill
%				\setlength{\parindent}{-1.4cm} 
%				%\hspace{1cm minus 0.25cm}
%				 \begin{subfigure}[b]{0.85\textwidth}
%           \centering
%				     
%				 \includegraphics[width=0.85\linewidth]{Line5.pdf}\vspace{-5pt}%Fig3--1bb
%				     \end{subfigure}
%				\indent
%	\caption{Number of tag reads as a function of splitting points for experiments and formulations in various address lengths and cache size and associativity.}\vspace{-13pt}		
%	\label{fig:5}
%\end{figure*}

 \begin{figure*}[h]
%\captionsetup{font=footnotesize}
	\centering
		\includegraphics[width=1\linewidth]{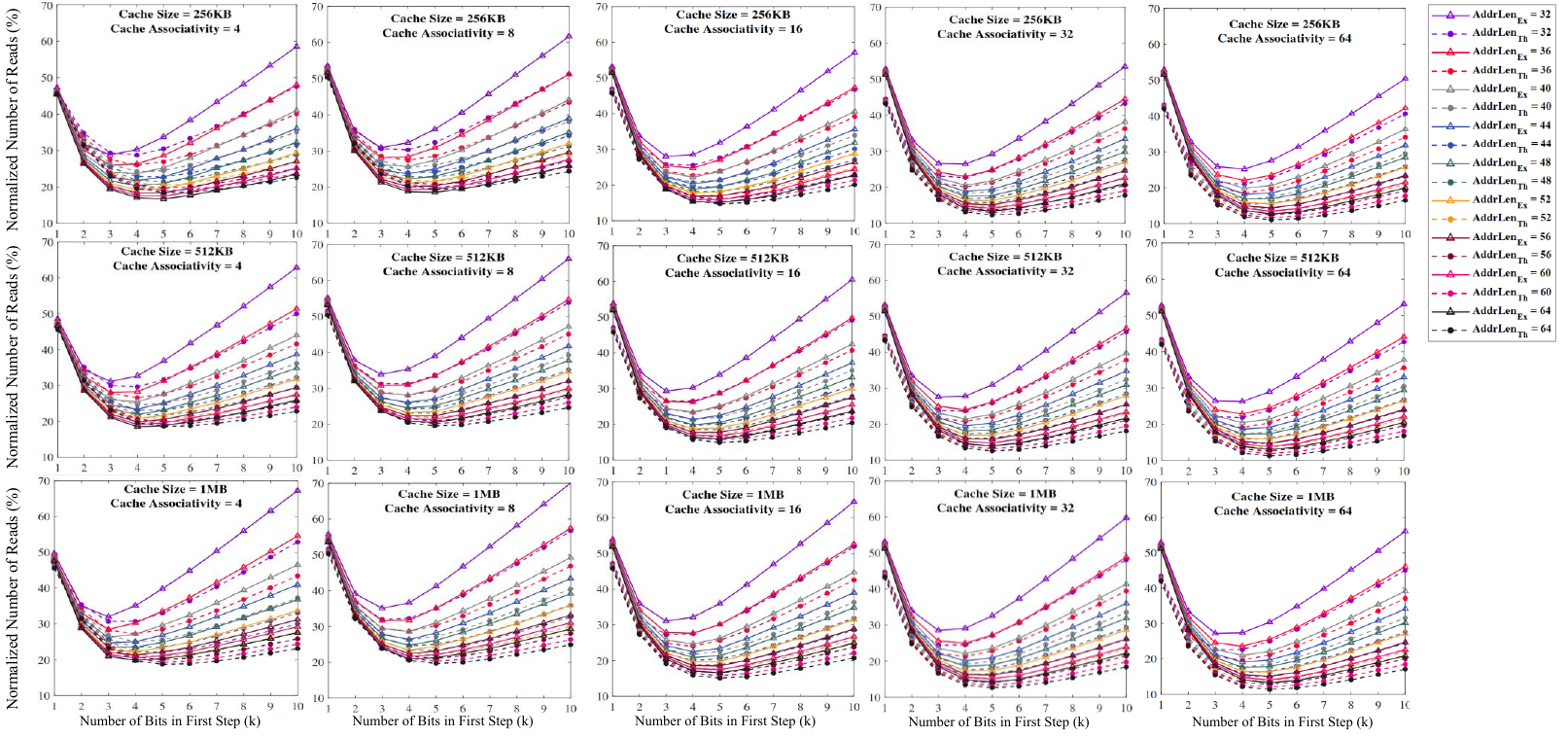}%2017
	\caption{Number of tag reads as a function of splitting points for experiments and formulations in various address lengths and cache size and associativity.}\vspace{-5pt}		
	\label{fig:5}	
\end{figure*}

%\begin{figure*}[t!]
%    \centering
%    \begin{subfigure}[b]{1\textwidth}
%           \centering
%           \includegraphics[width=0.9\linewidth]{my1.pdf}%surface1-HF.pdf
%    \end{subfigure}
%    \begin{subfigure}[b]{1\textwidth}
%            \centering
%          \includegraphics[width=0.9\linewidth]{my1.pdf}%surface2-HF.pdf
%    \end{subfigure}
%    \begin{subfigure}[b]{1\textwidth}
%            \centering
%          \includegraphics[width=0.9\linewidth]{my1.pdf}%surface2-HF.pdf
%    \end{subfigure}
%	\caption{Number of tag reads as a function of splitting points for experiments and formulations in various address lengths and cache size and associativity.}\vspace{-13pt}		
%	\label{fig:5}
%    \end{figure*}

\begin{figure*}[t!]
    \centering
    \begin{subfigure}[b]{1\textwidth}
           \centering
           \includegraphics[width=0.9\linewidth]{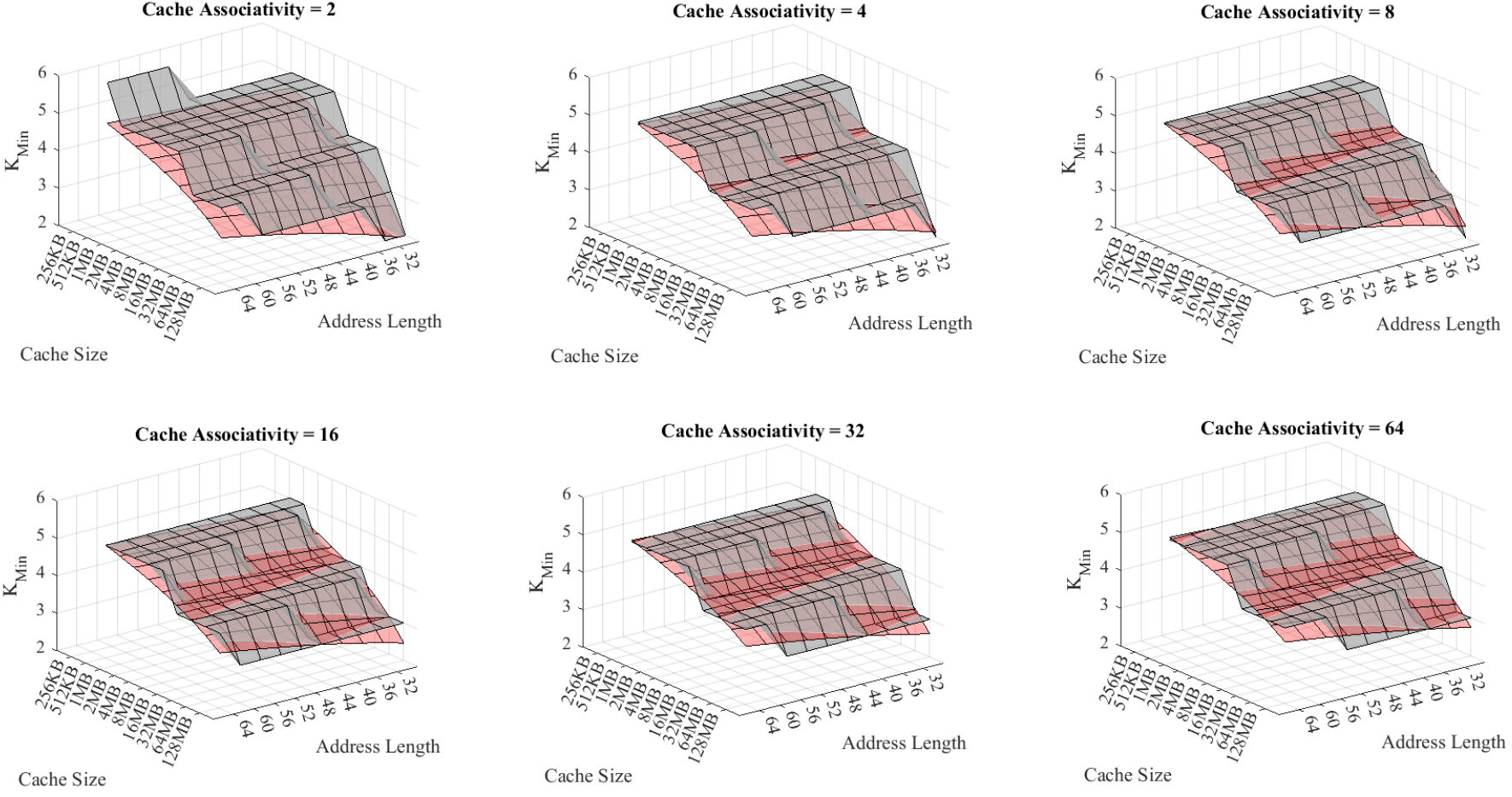}%surface1-HF.pdf
    \end{subfigure}
    \begin{subfigure}[b]{1\textwidth}
            \centering
          \includegraphics[width=0.9\linewidth]{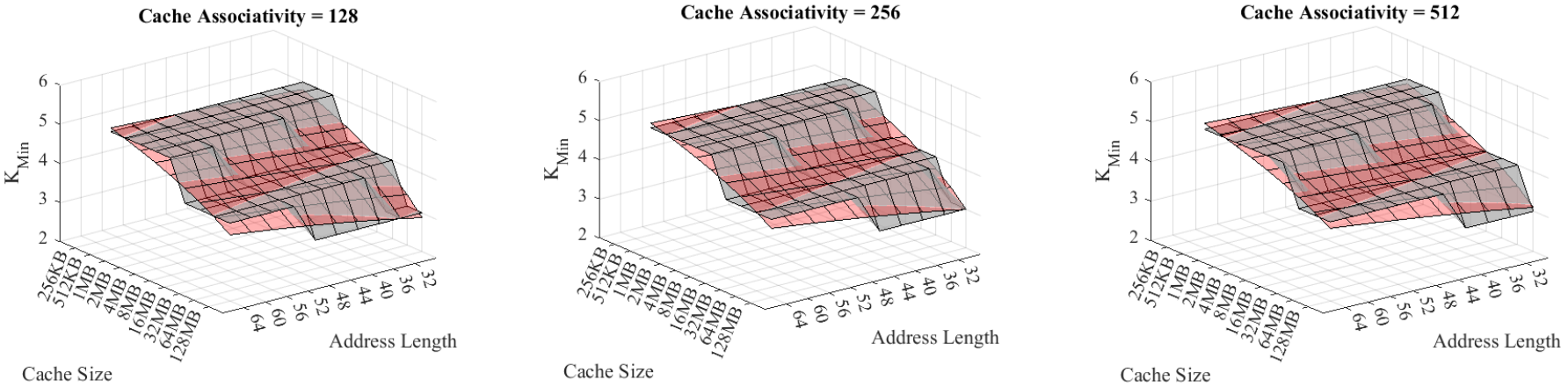}%surface2-HF.pdf
    \end{subfigure}
	\caption{Optimum splitting point ($K_{min}$) derived from Eq. (\ref{eq:1}), i.e., stairstep surfaces, and Eq. (\ref{eq:4}), i.e., smooth surfaces, for a wide range of cache sizes, associativities, and address lengths.}\vspace{-10pt}
				\label{fig:basics}
    \end{figure*}

The sixth subfigure in Fig. \ref{fig:workload-main} is the integration of all average curves in different associativities.
Three key observations on this subfigure are as follows.
First, the larger the cache associativity, the greater reduction in tag reads is observed; the top curve is for the 4-way cache, and the bottom curve is for the 64-way cache.
Second, the optimum splitting point is 4 for associativities 4, 8, 16, and 32, and it increases to 5 for associativity 64.
These points are highlighted in the curves.
This observation indicates that the optimum splitting point is directly affected by cache associativity, but a very large change in the latter slightly changes the former.
Third, considering the optimum splitting points, the left side of the curves is semi-exponentially descending, and the right side of the curves is semi-linearly ascending.
This observation indicates that for small/large values of splitting point (smaller/larger than the optimum), the reads are dominated by the tags in the second/first comparison step.
%\hl{The key observation of Fig. 4 is that the trend in the number of tag reads in various splitting points is almost similar in all workloads and is not largely affected by the workloads' behavior. Hence, the optimum splitting point is not meaningfully affected by the workload behavior, but by the cache configuration.} 

The normalized number of reads in the first step linearly depends on the splitting point and is independent of the cache associativity.
For large values of splitting points, the majority of tags are eliminated in the first step, which diminishes the contribution of the second step to the total reads.
For very small values of the splitting point, a large fraction of tags contributes in the second step, which exponentially decreases by a slight increase in the value of the splitting point.

%%shrink
%These observations are based on the fact that the number of reads in the first step of the tag comparison is linearly increased by increasing the tag-splitting point.
%This is obvious as this value per cache access is the multiplication of the cache associativity and the splitting point value.
%On the other hand, the number of reads in the second comparison step per cache access reduces by increasing the tag-splitting point in two aspects: 1) the length of the remaining tag bits linearly shrinks and 2) the number of tags remains for the second step is exponentially decreased.

\begin{figure*}[t]%\vspace{-5pt}
				\centering%\vspace{-1pt}
				%\vspace{-10pt}
				\includegraphics[width=1\linewidth]{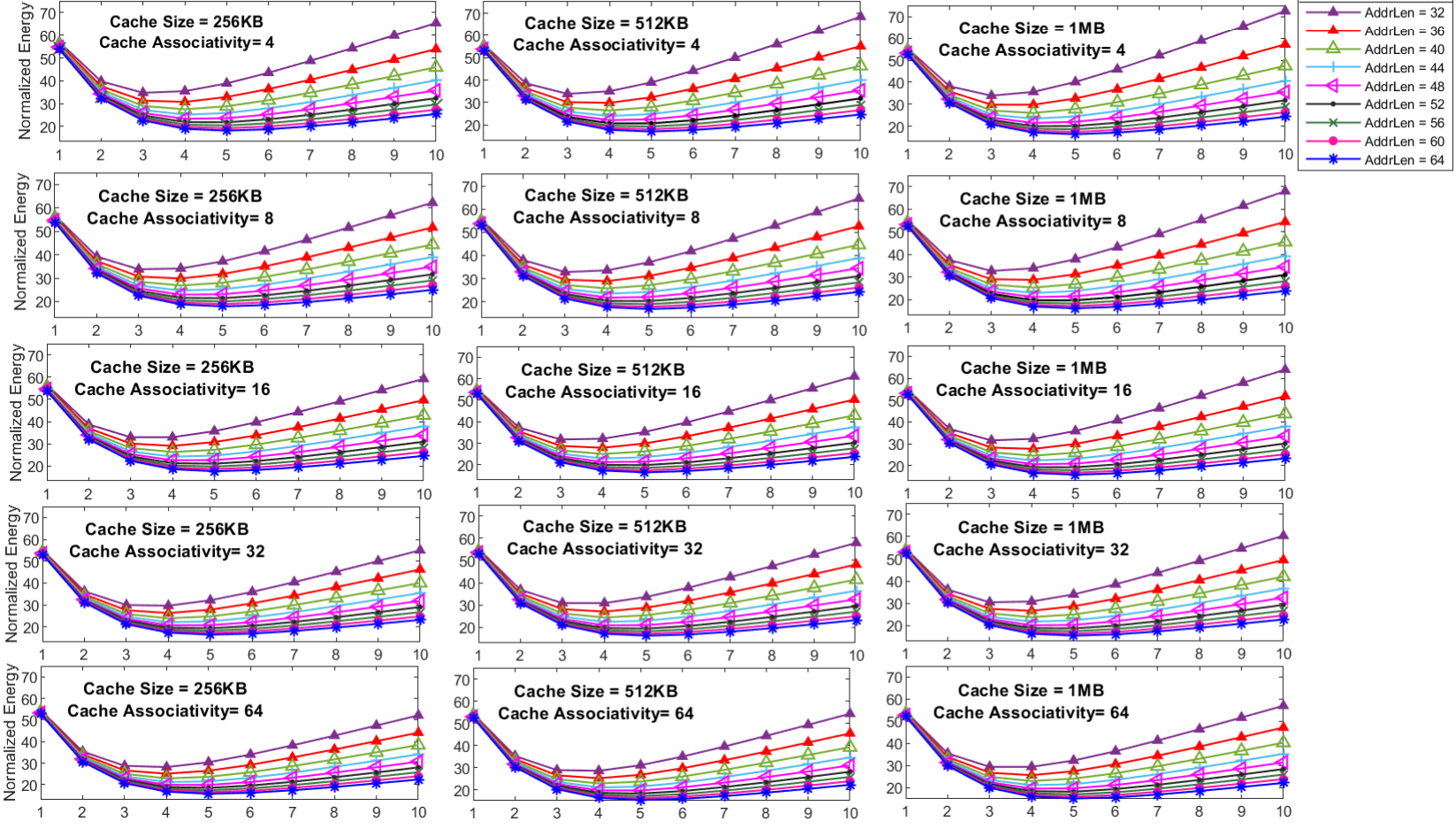}\vspace{-5pt}%table-1
				\caption{The effect of tag splitting point on energy consumption in various cache sizes, associativities, and address lengths.}
				\label{fig7}\vspace{-10pt}
\end{figure*}

\subsection{Comparison of Experiments and Formulations}
Fig. \ref{fig:5} provides a match between the experiments and formulations.
Considering the average curve of all workloads in each configuration as the representative of the experiments, their conformity to the formulated values is illustrated.
Three columns of subfigures in Fig. \ref{fig:5} depict the results for cache sizes of 256KB, 512KB, and 1MB, and five rows are related to cache associativities of 4 to 64.
Nine solid/dotted lines in each subfigure are for experiments/formulations considering the address lengths from 32-bit to 64-bit in the stride of 4.

Considering each subfigure in Fig. \ref{fig:5}, three main observations are evident.
First, the tag reads that reduction is greater for larger address lengths.
As an example, the normalized number of reads in the 256KB four-way cache (subfigure in upper-left corner) is about 30\% for an address length of 32-bit and is further decreased to 18\% for a 64-bit address.
The justification for this observation is that the tag partitioning approach reduces the number of reads by eliminating the tags from the second step of tag comparison and address length only affects the number of reads in the second step for a specific splitting point; address length has no effect on the number of bits reads in the first step of tag comparison, which is the common part of reads in the baseline and partitioned tag.
With the same reasoning, the gaps between curves of different address lengths get wider for larger splitting points since the effect of eliminating a tag from the second step of the comparison is higher in longer addresses.
Second, the optimum splitting point is almost the same in experimental and formulated curves.
Third, and more interestingly, better matching between experimental and formulated curves is observed around the optimum splitting point. By moving away from the optimum point, the pair of curves diverges.

%\begin{figure*}[t!]
%    \centering
%    \begin{subfigure}[b]{0.8\textwidth}
%           \centering
%           \includegraphics[width=1\linewidth]{surface1-HF.pdf}
%    \end{subfigure}
%    \begin{subfigure}[b]{0.8\textwidth}
%            \centering
%          \includegraphics[width=0.97\linewidth]{surface2-HF.pdf}
%    \end{subfigure}
%	\caption{STT-MRAM read and write operations: (a) reading `0', (b) reading `1', (c) writing `0', and (d) writing `1'.}\vspace{-10pt}
%				\label{fig:basics}
%    \end{figure*}

Considering each column of subfigures in Fig. \ref{fig:5} depicting the results for different cache associativities in specific cache sizes, two key phenomena are observed.
First, a higher matching between pairs of curves, i.e., experiments and formulations, is observed for smaller cache associativities.
Note that even in the worst-case matching, the optimum splitting point is still the same in pairs of curves.
Second, the reduction in the number of reads is larger in higher associativities for all cache sizes and all address lengths.
This is in accordance with the observation already discussed in Fig. \ref{fig:workload-main}.

\begin{figure*}[t]%\vspace{-5pt}
				\centering%\vspace{-1pt}
				%\vspace{-10pt}
				\includegraphics[width=0.9\linewidth]{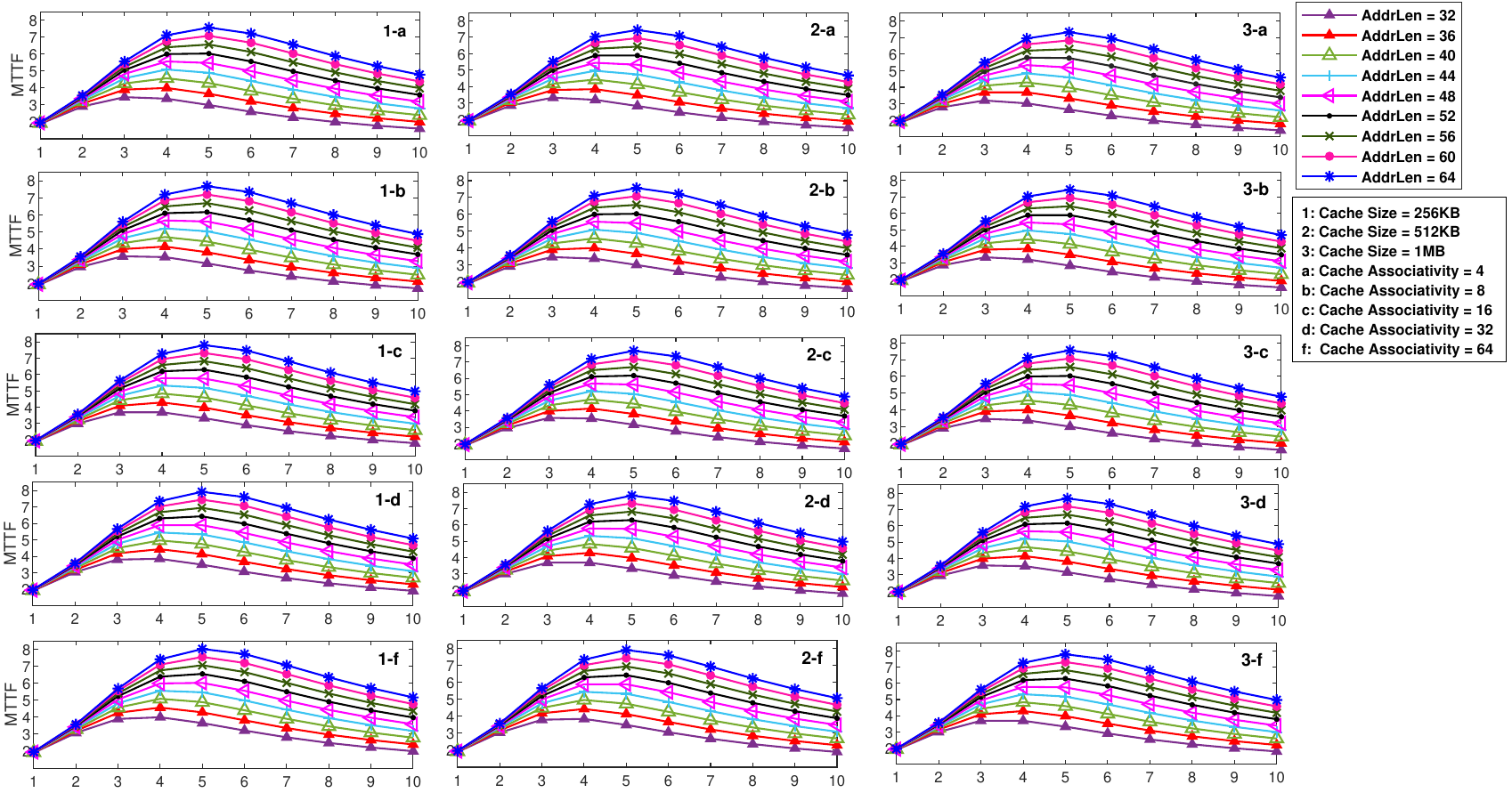}\vspace{-5pt}%table-1
				\caption{The effect of tag splitting point on mean-time-to-failure (MTTF) of tag array in various cache sizes, associativities, address lengths.}
				\label{fig8}%\vspace{-10pt}
\end{figure*}

\subsection{Sensitivity of Optimum Point to Cache Configuration}
After justifying the agreement of formulations with the experiments for different cache configurations, Fig. \ref{fig:basics} is constructed to deeper investigate how the optimum splitting point is theoretically affected by the cache configuration parameters for a very wide range of configuration values.
Considering each subfigure for specific cache associativity, there are two surfaces, a smooth one and a stairstep one.
The stairstep surface represents optimum splitting points derived from Eq. (\ref{eq:3}) (calculates the number of reads for various splitting points).
This is done by finding $k$ with minimum output, which practically must be an integer value. The smooth surface represents the optimum splitting points calculated using Eq. (\ref{eq:4}), i.e., the point at which the derivative in Eq. (\ref{eq:3}) is zero.
Theoretically, this point could have any real value.
The x-axis in Fig. \ref{fig:basics} is the cache size varying from 256KB to 128MB, and the y-axis is the address length ranging from 32-bit to 64-bit.
The nine subfigures illustrate the results for cache associativities from 2 to 512.

Considering Fig. \ref{fig:basics}, the following observations are worthy of attention.
First, it can be observed for each subfigure that the stairstep surface derived from Eq. (\ref{eq:3}) is equivalent to the round function of the smooth surface.
Second, this is valid for all subfigures in that the optimum splitting point gets smaller by increasing the cache size and gets larger by increasing the address length.
Considering Eq. (\ref{eq:3}), which is a function of splitting point, tag length, and associativity, both cache size and address length only affect the tag length.
Tag length gets smaller by increasing the cache size and/or decreasing the address length.
Both of these changes lead to an increase in the optimum splitting point, which can also be expected from Eq. (\ref{eq:4}).
Third, the sensitivity of the optimum splitting point to cache size and address length diminishes in larger cache associativities.
As is evident, the surfaces become flatter in higher-associative caches.

In summary, our evaluations demonstrate a very close match between the formulations and the experiments, which is a justification for the formulations.
Considering a very wide range of cache configurations, the observations show that the optimum splitting point varies from 2 in a 2-way associative 256KB cache with an address length of 64-bit to 6 in a 2-way associative 128MB cache with an address length of 32-bit.
Therefore, for today’s realistic cache configurations, the optimum splitting point resides between 3 and 5.
The interesting fact for this range is that, as observed in Fig. \ref{fig:5}, the final reduction in the number of reads is very close to the optimum point and its neighboring values.
Therefore, considering the safe range of [3, 5], it can be guaranteed that a very close to maximum read reduction is achieved. 

\subsection{Energy Consumption}
Fig. \ref{fig7} depicts the effect of read reduction on the SRAM-based tag energy consumption for various cache configurations. While a small fraction of tag energy consumption is independent of the number of bits read, e.g., set indexing, its majority is affected by tag partitioning, and hence, by the proper selection of the splitting point. Compared to the baseline, the trend in energy reduction corresponds to that of the reduction in the number of reads. Again, a higher reduction is observed for longer address lengths and higher associativities. Considering the 4-way associative 256KB cache, the maximum energy consumption reduction (on the optimum splitting point) is 66\% for 32-bit address length and gradually increases for longer addresses up to 82\% for 64-bit address length. 
Energy consumption reduction in the 32-way associative cache with a size of 256KB is 71\% and 84\% in address lengths of 32- and 64-bit, respectively. 
On the other hand, if not properly selecting the tag splitting point, the tag energy consumption can increase as high as twice the optimum point in some configurations.

\begin{table}[t]\vspace{-5pt}
				\centering%\vspace{-1pt}
				\caption{Details of STT-MRAM cells parameters.}\vspace{-5pt}
				\includegraphics[width=0.95\linewidth]{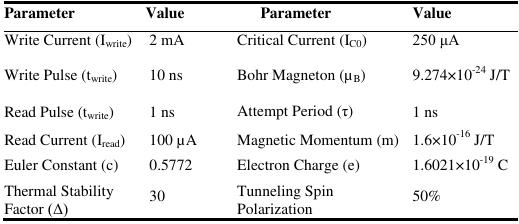}\vspace{-5pt}%table-1
				\label{ft}\vspace{-10pt}
\end{table}

\subsection{Reliability Analysis}
In addition to energy consumption reduction, tag partitioning reduces the error rate of STT-MRAM-based tag arrays. As the metric of tag reliability, we report the \textit{Mean-Time-To-Failure} (MTTF) of the tag array using the framework presented in \cite{cheshmi}. 
There are three sources of errors in STT-MRAM tag, i.e., read disturbance, retention failure, and write failure, all of them affected by process variation. Considering the 32nm technology node, device- and circuit-level parameters of STT-MRAM cache for MTJ thermal stability factor and read current, and Gaussian distribution of process variation, Table \ref{ft} reports the related cache parameters and per-cell error rates. Considering three sources of errors, read disturbance is the main source of unreliability in the tag array due to its frequent read accesses. On the other hand, tag partitioning mainly affects read disturbance rate by reducing the number of reads, which has a direct effect on error rate reduction. To calculate the MTTF, we take into account the total probability of error occurrence per tag cell during read operations. During the workload execution, all read accesses to the tag cell have been considered, and the cache reliability for a workload is defined as the probability of error-free data read for all tag accesses. Given the reliability and execution time, and assuming exponential distribution for reliability with error rate parameter $\lambda$, the MTTF for each workload is the inverse of $\lambda$.

Fig. \ref{fig8} illustrates the MTTF of the tag array considering various cache configurations and tag splitting points, normalized to the baseline. The observations indicate that higher MTTF enhancement is achieved in the longer address lengths and higher cache associativities. MTTF is enhanced by 3.4x and 7.8x in a 4-way associative 256KB cache in address lengths of 32- and 64-bit on the optimum splitting point of 3 and 5, respectively.  On the other hand, the proper selection of the tag splitting point has again a significant effect on the MTTF improvement by tag partitioning. Considering a single cache configuration, the blue curve in subfigure 2-d, for example, the MTTF enhancement is reduced from 7.9x to 5.6x when the splitting point is moved from the optimum point of 5 to an unsuitable point of 9.

\begin{table}[t]\vspace{-5pt}
				\centering%\vspace{-1pt}
				\caption{Summary of the tag splitting point value and selection method in previous work.}%\vspace{-10pt}
				\includegraphics[width=0.8\linewidth]{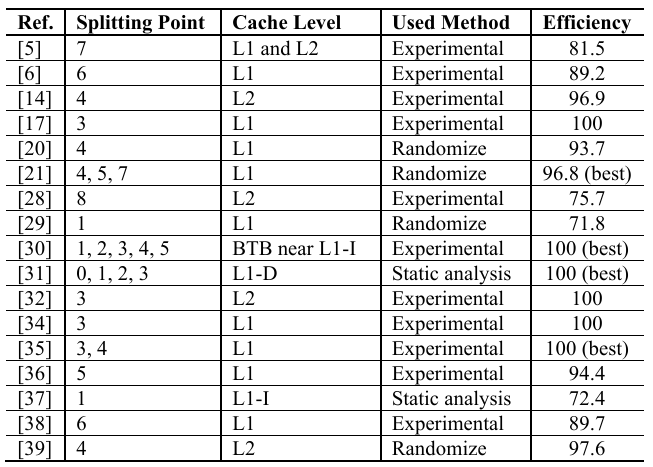}\vspace{-5pt}%table-1
				\label{jadid}\vspace{-10pt}
\end{table}

\begin{table*}[t]%\vspace{-5pt}
				\centering%\vspace{-1pt}
				\caption{Timing details of data and tag arrays for a 256KB cache in various address lengths.}\vspace{-10pt}
				\includegraphics[width=0.9\linewidth]{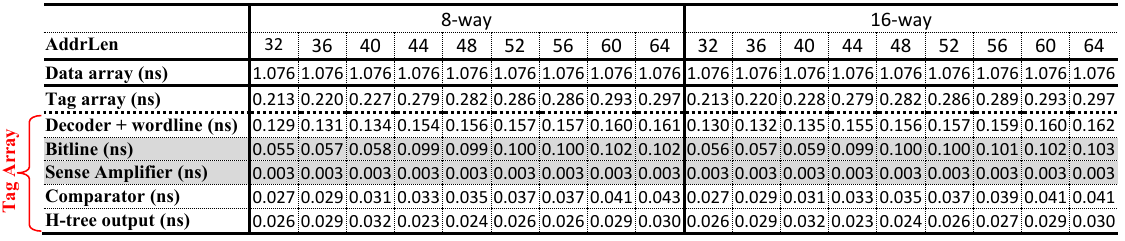}\vspace{-5pt}%table-1
				\label{delay}\vspace{-10pt}
\end{table*}

In summary, the evaluations confirm the alignment of the formulations with the experiments, enabling a theoretical determination of the optimal tag splitting point with high accuracy during the early design stage. These formulations eliminate the need for exhaustive searches to find a suitable splitting point or for its random selection, which can compromise efficiency. Our observations indicate that the optimal splitting point is not significantly influenced by workload behavior due to the randomness in the combination of lower order bits of tags in a cache set, but rather by the cache configuration. Table \ref{jadid} presents details of prior work on tag partitioning, including their selected splitting point,  cache level, and the method of acquisition.
Due to the lack of a reliable formulation, some methods have identified the appropriate point by examining various cases, while others have considered a random point that was sometimes fortuitously close to the optimal point.

Tag splitting necessitates minor modifications to the cache controller and tag wiring with negligible area and energy overheads to achieve significant energy savings. However, accessing the tag array in two steps does increase tag access latency. This increase in latency does not affect cache access time as long as the total tag operations conclude before the data array access operation is completed. As reported in \cite{3rset}, the latency of tag operations, in both conventional and splitting configurations, is lower than the data array latency across various cache sizes and associativities. 
Hence, the increase in tag splitting point has no impact on read cycles, overall cache latency, or the energy consumption of other cache components. Table \ref{delay} illustrates the timing parameters of a 256KB L2 cache for 8- and 16-way associativities and different address lengths. Tag partitioning doubles the delays related to the two highlighted rows since the tag bits are accessed and compared in two phases. Because the access time of the data array is significantly greater than that of the tag array, this increase in tag array delay does not affect cache performance.

\section{Conclusion}\label{sec:conclusion}
Tag array in set-associative on-chip cache memory contributes to a large fraction of energy consumption and vulnerability to errors because of very frequent and parallel read accesses. 
Tag partitioning as a well-known approach has been utilized in several studies to reduce the number of reads from the tag array by splitting the tag comparison operation into two steps, in which very few lower order tag bits of all cache ways in the target set are activated in the first step and only the partially matched ways are activated in the second comparison step.
Although the value of the tag-splitting point is the key factor in the efficiency of the tag partitioning-based scheme, there is no investigation on the proper selection of this parameter and its effects in the existing studies.
This paper formulated the tag partitioning efficiency to find the optimum point of tag splitting and conducted comprehensive analysis and experiments for a wide range of system configurations to validate the accuracy of the proposed formulation.
This formulation makes the designers and researchers capable of determining the optimum tag-splitting point at the early design stage and precisely estimating the tag partitioning efficiency for energy consumption and error rate reduction.

\bibliographystyle{IEEEtr}
\bibliography{bibliography}
%%%%%%%%%%%%%%%%%%%%%%%%%%%%%%%%%%%%

%\newpage

\begin{IEEEbiography}[{\includegraphics[width=1in,height=1.25in,clip,keepaspectratio]{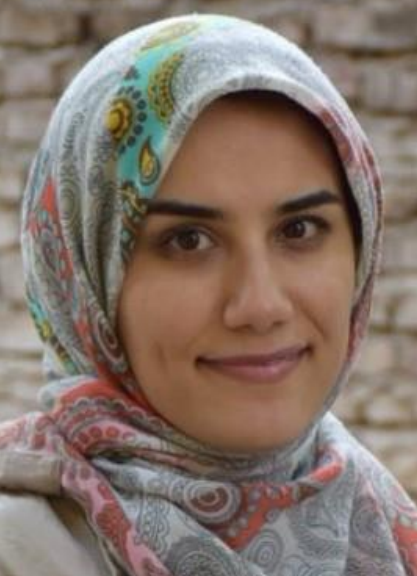}}]{Elham Cheshmikhani}
is an assistant professor at the Department of Computer Science and Engineering at Shahid Beheshti University, Tehran, Iran. She also served as an assistant professor at the Department of Computer Engineering at Amirkabir University of Technology (Tehran Polytechnic), Tehran, Iran, from September 2022 to September 2023. Her educational background includes a B.Sc., M.Sc., and Ph.D. degrees in Computer Engineering from the Iran University of Science and Technology (IUST), Amirkabir University of Technology (Tehran Polytechnic), and Sharif University of Technology (SUT) respectively, obtained in the years 2011, 2013, and 2020.
From January 2021 to September 2022, she worked as a postdoctoral researcher at the Department of Information Engineering and Mathematics at the University of Siena, Siena, Italy. During her postdoctoral research, she collaborated with Huawei R\&D (UK) Ltd. She was a member of the Design and Analysis of Dependable Systems (DADS) at Tehran Polytechnic from 2011 to 2015 and a member of the Dependable Systems Laboratory (DSL) at SUT from 2016 to 2018, and the Data Storage, Networks \& Processing Laboratory (DSN) at SUT from 2018 to 2021. Her research interests encompass emerging nonvolatile memory technologies, processing-in-memory, RISC-V ISA design, hardware accelerators, SoC design, dependable systems design, and storage systems. Notably, she achieved the Best Paper Award at IEEE/ACM Design, Automation, and Test in Europe (DATE) in 2019.

\end{IEEEbiography}


\begin{IEEEbiography}[{\includegraphics[width=1in,height=1.25in,clip,keepaspectratio]{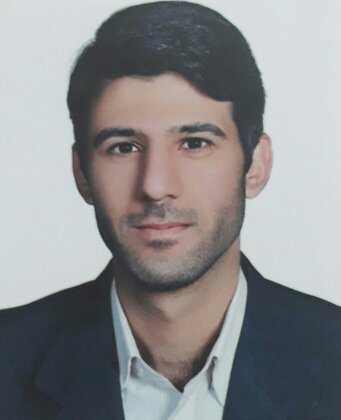}}]{Hamed Farbeh}
received the B.Sc., M.Sc., and Ph.D. degrees in computer engineering from Sharif University of Technology (SUT), Tehran, Iran, in 2009, 2011, and 2017, respectively. He was a member of the Dependable Systems Laboratory (DSL) at SUT from 2007 to 2017, was with the Embedded Computing Laboratory (ECL), KAIST, Daejeon, South Korea, as a Visiting Researcher from October 2014 to May 2015, and collaborated with the Institute of Research for Fundamental Sciences (IPM), Tehran, Iran, as Postdoc fellow from May 2017 to January 2018.
He is currently a faculty member of the Department of Computer Engineering, Amirkabir University of Technology (Tehran Polytechnic-AUT), Tehran, Iran, where he established the Intelligent Computing and Communication Infrastructure Laboratory (ICCI) and is the director of Computer Systems Architecture and Networks (CSAN) group. He is also a member of the board of the Cyber-Physical Systems Society of Iran (CPSSI) since 2018. His current research interests include reliable memory hierarchy, emerging memory technologies, AI processors, and cyber-physical systems.
\end{IEEEbiography}

%\vfill

\end{document}